# Open data quality

Anastasija Nikiforova[0000-0002-0532-3488]

University of Latvia, 19 Raina Blvd., Riga, LV-1586, Latvia
Nikiforova.Anastasija@gmail.com

**Abstract.** The research discusses how (open) data quality could be described, what should be considered developing a data quality management solution and how it could be applied to open data to check its quality. The proposed approach focuses on development of data quality specification which can be executed to get data quality evaluation results, find errors in data and possible problems which must be solved. The proposed approach is applied to several open data sets to evaluate their quality. Open data is very popular, free available for every stakeholder - it is often used to make business decisions. It is important to be sure that this data is trustable and error-free as its quality problems can lead to huge losses.

## 1 Introduction and Motivation

The data quality problem is topical since over 50 years, and many different approaches are discussed in scientific publications addressing data quality issues [1].

In the major part of sources, in the center of attention is informal definition of data quality characteristics and their values measuring but mechanisms for specifying data quality characteristics in formalized languages usually are not considered. In previous research [1], [2] author's colleagues have already described the main points of suggested approach which could solve data quality problems. The main task of this research is to provide short overview on this mechanism and to apply it to open data to check its quality. Open data is often used to make important decisions according to results of its analysis. Its quality has huge impact on making decisions. Open data is usually used with assumption that it is of high quality and is ready to process without additional activities (such as quality checks) but this assumption hasn't been approved.

The main aim of this paper is to analyze open data quality, checking assumption that open data is of high quality. Secondary aim is to apply in [1] proposed approach to real data sets to check its appropriateness to the specified task.

The paper deals with following issues: overview on the related research (Section 2), problem statement (Section 3), a description of the proposed solution (Section 4) open data analysis (Section 5), results (Section 6), evaluation plan (Section 7).

## 2 State of the Art

One of the main theories in data quality area is TDQM (an overview of existing methodologies is available in [3]). As the most popular way to evaluate data quality is to use data quality dimensions [4], [5], Data Management Association International UK Working Group defined 6 data quality dimensions: completeness, uniqueness, timeliness, validity, accuracy, consistency [1]. This approach is widely used and most of the existing studies focuses on data quality dimensions [6], [7] and their application to data sets. [6] proposes to analyze data sets according to specific dimensions (traceability, completeness, compliance) calculating the metrics with a RapidMiner. Unfortunately, definition of specific requirements for specific fields isn't provided. Some studies, for example, [8], [9] defines guidelines for data holders and open data portals which could check and help to improve data quality before it will be published.

Most of the solutions (frameworks) examine open data portals. For example, [10] and [11] provides overview of automated quality assessment framework which allows to discover and measure quality and heterogeneity issues in open data portals. It can be used by portals holders and inspectors to examine the whole portal quality (but not several open data sets). Another example is the PoDQA project [12] - a data quality model for Web portals based on more than 30 data quality attributes. Such high number of attributes can lead to misunderstandings as some criteria are similar and sometimes the difference between them is minimal (possibly, DAMA classification could be more appropriate). These solutions also aren't appropriate to examine specific data sets.

Other research mainly deals with linked data quality [13], [14], [7], [15], mainly focusing on DBpedia and RDF triples (researching the Semantic Web or converting data to RDF). These studies are extremely popular and useful to check linked data quality but aren't appropriate to check standalone data set quality by non-IT expert.

[9] solution can be applied to specific data sets and provides graphical results but analysis supposed to be done involving IT-experts - users can specify quality dimensions which should be included in analysis of data sets. Moreover, developers of this solution say that it is not able to check correctness of specific data formats. It also isn't known if there are available protocols which could be used to identify specific records which must be revised and which quality must be improved.

Probably the most widely known data quality tools are Microsoft SQL Server Integration Services (SSIS) and Data Quality Services (DQS). SSIS offers wide range of features for data migration and designing of ETL processes [2], [16], [17]. User can define how the process should flow and perform some task on different interval but its memory usage is high and it conflicts with SQL [18], [19]. SSIS wasn't designed specifically for data quality management and there are no known its uses for data quality management which were not related to data migration. DQS is knowledge-driven data quality tool (SQL Server component) which is designed for data quality analysis and improvement. Its main functional components are: knowledge base; data matching; data cleansing; data profiling [20]. DQS has several disadvantages: possibility to analyze only one table per time; domain formats limitations; minimal data matching threshold is 80%; high resource usage analyzing bigger data amounts (CPU, Disk and Memory usage grows up to 100%).

To sum up, most of the existing solutions are unsuitable for industry professionals without appropriate IT background as their aim is to check data quality (in general). There are tools and theoretical research which can be used to analyze and improve data quality, but they have their own limitations and disadvantages. Definition of data quality dimensions and methods for its quantitative evaluation is one of the most important step was ever made. Mechanisms for data quality characteristics specification in formalized languages such as [1] (developed by author's colleagues) which could provide users with possibility to easily check specific data sets quality, defining specific requirements for each column which depends on the purpose for which data will be used, weren't proposed. Existing research concluded that open data has data quality issues. It is important to explore it and offer approaches which could improve it.

## 3 Problem Statement

As stated in the previous section, most of the existing solutions are unsuitable for professionals without appropriate IT background. Moreover, the most of research can't be applied to the specific data sets, defining data quality requirements for specific columns. The aim of [1] approach is to provide every user (even users without appropriate IT background) with possibility to check data quality, letting define specific requirements to every column.

Traditional quality checks implementation is hard to test and to manage. It is one of the reason why data quality is rarely checked and data often contain data quality defects. To solve a given problem, they are substituted with universal decision – quality checks are separated from program code. As data quality has relative and dynamic nature - data quality requirements depend on data application and staged data accruing, following data quality control system main principles can be defined: (1) data quality requirements must be formulated on several levels – for specific data object, data object in the scope of its attributes, data object in the scope of database and data object in the scope of many databases; (2) at the first stage, data quality definition language must be graphical DSL, which syntax and semantics can be easily applied to any new information system. It must be simple enough to understand to provide industries specialists possibility to define data quality requirements without IT specialists supervision; (3) data quality should be checked in various stages of the data processing, each time using its own individual data quality requirements description; (4) as data quality requirements depend on data application, quality requirements must be formulated in platform independent concepts avoiding checking inclusion in information systems source code.

Each specific application can have its own specific data quality checks - it is important to develop platform which (1) will allow to define quality requirements and (2) will be possible to configure to get specific DSL. It is important to achieve data quality requirements specification execution where data quality defects could be found in the result of its execution.

Nowadays open data is extremely popular and there are a lot of portals which provide users with a wide range of different data such as company registers, education and culture, civilians, development, environment and other. One of the main open data portals of Latvia is [21]. Open data key idea is to collect data which will be free available to everyone to use, analyse and republish as they need without any restrictions. 8 principles which open data must satisfy are: data must be complete, primary, timely, accessible, machine processable, non-discriminatory, non-proprietary and licence-free [21]. These principles don't focus on data quality and it let make assumption that open data may have quality defects. As this data is used to make important business decisions, it is important to be sure that data is of high quality and is trustable. This data is used without additional activities (such as quality checks) making assumption that it has been already checked by data holders and is ready for further processing.

Author of this paper suppose that (1) open data often isn't checked before it is published and (2) its analysis will identify quality problems. This paper summarizes analysis of open data quality. Given analysis is made using previously described approach.

## 4 Research Method

According to TDQM methodology [22], data quality lifecycle can be described by four interconnected data quality control phases: (1) definition, (2) evaluation, (3) analysis, (4) improvement. This research deals with steps 1-3 but data quality improvement isn't deeply researched as it can be ensured with previously mentioned MS DQS.

Data quality control process is given as a cycle as all phases must be systematically repeated to achieve and to keep high data quality. Efficiency of this mechanism can be achieved by repetition of 2$^{nd}$ and 3$^{rd}$ phases. Moreover, data in data storages is continuously changing and new data can bring new data quality problems or cause new requirements. Used approach [1] suppose that assessment criteria can be changed, and new data quality requirements or different data quality conditions can be defined at the new cycle iteration.

Data quality control system which was proposed by this paper authors colleagues in [1] consists of 3 main components: (1) data object, (2) quality requirements, (3) description of quality measuring process. These components form data quality specification where data object description define data which quality must be examined. Quality requirements defines conditions which must be fulfilled to admit data as qualitative (nullability, format, data type etc.). Description of quality measuring process define procedure which must be performed to assess data quality. In the centre of data quality control system is data object. Data object is a set of existing object parameters. It has only those data which must be analysed. It reduces amount of processed data (kind of denormalization). According to structure of the data objects class, data object has random number of fields, and their attributes prescribe possible value constraints and other data object classes. Data objects classes are necessary to define data quality requirements for collections of data objects, for example, for database which stores data about people, invalid person data amount can't exceed 1% of all records - such quality dimension is measurable only if all parameters' values of specific record are checked one by one by, relating number of errors to the number of all processed records.

It is recommended to create data quality model which consists of graphic models, where each diagram describe specific part of data quality check. All checks of one business process step are unified in the packages. All packages form data quality model. Diagrams consists of vertexes which have data quality control activities and are connected with arcs which specify order of these activities. It can be used in two ways: (1) informal which has description of necessary checks activities in natural language where diagram symbols have textual description of activities and (2) executable which can be get by conversion of informal model substituting informal texts by code, SQL queries or another executable object.

In this research, all these components are described with language metamodels - each component is provided with its graphical representation. 3 language families where defined. Their syntax is defined by representational models supported by tools developing platform DIMOD which allow to define a variety of DSL with different data objects structures and after its inserting in repository, DIMOD behaves as DSL graphic editor. Involved languages are developed as graphic languages and are related with tools possibilities of the development platform DIMOD which is advised to be used instead of using specific data objects definition language [1].

## 5 Open Data analysis

This section demonstrates proposed idea to check open data quality. This paper shows the analysis of 3 data sets provided by Municipal of Riga [21] and summarizes results of analysis of The Register of Enterprises of the Republic of Latvia [23]. Author assumes that as these portals are main open data portals in Latvia, they should have data of higher data quality. Chosen data sets satisfy open data principles (see Section 2).

Municipal of Riga provides data about educational licenses for 2013, 2014, 2015. Data is available in two formats - .xls (Microsoft Excel) and .csv (Comma Separated values) and consists of 9 columns: license requester, registration number, realized program, program type, realization place, hours, decision, terms, license number. All data format is varchar but in the result of data exploration (analyzing each column), more appropriate data formats, nullability constraints and other data quality requirements for each field were defined. As all analyzed data sets have the same structure, Figures 1-3 can be applied to all of them (making appropriate changes).

Figure 1 describes data in its original state (input data) in informal way.

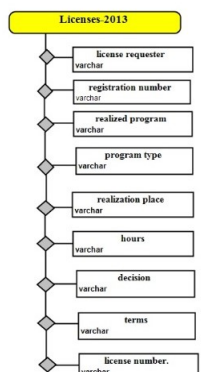 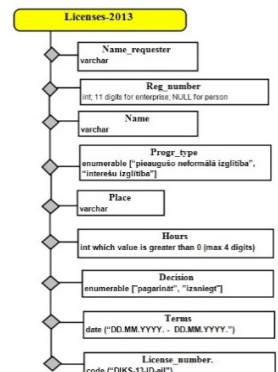

Fig. 1. Data object "Licences-2013" (input)     Fig. 2. Data object (database)

Figure 2 shows data state in database with value attributes and data quality specification which is used in data quality requirements specification.

When data quality requirements are inspected, data object fields attributes can be changed to attributes which are shown on Figure 3. At this stage of the research, analyzed data sets were imported to SQL Server database and all records were processed with SQL queries according to defined quality requirements and developed diagrams. In the future, this process will be automatized making diagrams executable.

4th data set is provided by [23] and has 22 columns - 396 952 records (proposed ideas can be also applied to larger data sets). As data object and quality specification are defined according to previously described principles, corresponding figures are not provided in this paper.

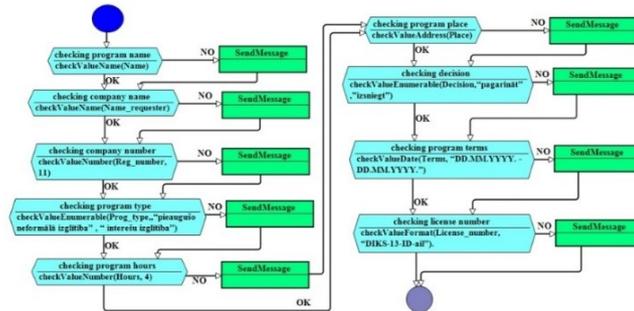

**Fig. 3.** Data object "Licences-2013" quality specification

The results of data processing are available in the next section.

## 6    Results

This paper covers 4 data sets analysis. These data sets were imported to SQL Server database and processed according to the proposed approach – defining data object which quality should be assessed and data quality requirements which data must fulfil (records were processed with SQL queries according to defined quality requirements). Each component is provided with its graphical representation.

1st – 3rd processed data sets provided by Municipal of Riga [21] don't have any significant quality weaknesses. Only one field was partly filled – "stundas" (hours) but as it was observed in all data sets (446 of 501 records have NULL values), author suppose this field can had NULL values. Other fields which must have NOT NULL values had values which corresponds to data quality specification. Author concluded that given data sets are of high quality and can be used to make analysis and business decisions.

Results of analysis of the 4th data set [23] are very unexpectable – it has many quality problems (11 of 22 fields have defects), for instance: 10 enterprises don't have name (error rate is 0.0025%); 94 records (0.024%) don't have registration date; 366 records (0.09%) don't have "Address"; 1 403 (0.35%) records don't have "Type_text" ( "Type" values full name which are NOT NULL); 20 496 (5.16%) records don't have index and 2 indexes have incorrect format; 1.15% of records don't have ATV code and 0.24% values format is incorrect; 646 (0.18%) enterprises have "terminated" date but don't have "closed" value - both values must be filled or empty). These results show that processed data set has quality problems even in those fields which have primary company information – name, registration number and registration date. It means that not all companies could be identified and contacted (as address field has NULL values). Some of these records has inconsistent data (active companies have date of termination). Moreover, there is no confidence that these companies exist (possibly, these records were added as test data, or they are too old but nobody has delete them).

Same approach was applied to another 3 Company Registers of other European countries and results of these checks were similar with previously stated – all data sets have at least few quality problems. Most of the quality problems are identified in (1) address fields - NULL values and (2) date fields - dubious date values, for example, there are companies which according to the registers were organised in 1552. As the number of detected data quality problems is quite low, these quality problems could be easily solved using proposed approach to identify them. It would significantly improve overall data quality.

It should be noted that in cases when number of empty values is quite big, there could be specific conditions which allow to leave these fields blank – but it is important result as it shows that proposed approach helps to detect such anomalies.

Main results of these analysis are that (1) proposed approach lets analyse data quality of data sets without knowledges about how this data was collected and processed by data holders – it is possible to analyse "foreign" data (it is huge advantage of proposed approach), (2) detect quality problems and anomalies in open data which is available for every stakeholder and can be used to make important business decisions, (3) open data often has data quality problems which must be detected before data will be used in analysis where data quality weaknesses could lead to huge losses.

## 7    Evaluation/Validation Plan

As it was previously stated in [1], data quality management mechanisms should be able to execute data quality specifications. To check results which were provided in the previous section, it is planned to achieve initial research aim – to make quality model executable involving interpreter which will (1) translate defined diagrams and (2) execute data quality checks, resulting records which have errors. When this step will be completed, it will be possible to analyze open data automatically, using the data quality model for data quality measurement. Currently, analysis was made manually - writing SQL queries, where SELECT part defines data object and WHERE part specifies data quality requirements, executing them one by one. It is supposed that results will be the same with results shown in this paper but process of quality checks (1) will be easier and faster (as there won't be necessity to write

SQL queries) and (2) could be used even by non-IT experts providing industries specialists with possibility to define data quality requirements with minimal IT specialists involving.

Results of analysis will be given to data holders letting to review them and to solve identified quality problems. Some of detected problems – "anomalies" (see Section 6) could be ignored as defined quality requirements is authors interpretation of stored data.

Proposed approach will be applied to more open data sets to make more general conclusions on open data quality. As it was mentioned in Section 5, this approach was already applied to another 3 data sets, collecting open data from different countries to make generalization on total open data quality. Future analysis will cover specific topic - Company Registers choosing specific use-cases as data quality depends on them.

In the future, given approach could be applied to Linked Data as it is important to check how data are interlinked.

## 8    Conclusions

The paper is a continuation of author' research in the area of executable models and DSL [1], [2], [16], [24] applying developed approach to open data to evaluate its quality. Open data analysis results draw the following conclusions:
- Proposed approach works and allows (1) to define data objects which quality will be analyzed retrieving them from different sources and (2) to make data quality assessment detecting data quality problems. Moreover, proposed approach can help to find anomalies in analysed data. It could be considered as new step in data quality management. It allows to analyse data without any limitations such as format or data amount – it is universal for many purposes;
- Open data is published for many years and this research proves that open data may have data quality problems (even more problems that were supposed) as there are no centralized checking of open data quality and (2) there is not enough research on open data quality. Stakeholders who use data making business decisions should know it and take in to account as it could have huge impact on their business.
- Open data quality depends on data provider but even trusted open data sources (as Company Registers) may have data quality problems.
- Open data quality must be inspected because, as it was shown in this research, it can have data quality problems which must be detected, explored and solved. Proposed approach is one of the most appropriate options which could be used to improve data quality even by non-IT experts.

This research must be continued to automatize proposed approach involving interpreter which will make data quality checking process easier and usable even by non-IT experts which could lead to global data quality improvements.

## Acknowledgments

This work has been supported by University of Latvia project AAP2016/B032 "Innovative information technologies".